\documentclass[aps,prl,singlecolumn,showpacs,superscriptaddress,groupedaddress]{revtex4-1}

\newcommand{\ket}[1]{\left\vert#1\right\rangle}

\newcommand{\expect}[1]{\left\langle#1\right\rangle}

\usepackage{pdfpages}
\usepackage{graphicx}
\usepackage{epstopdf}
\usepackage{setspace}
\usepackage{bbold}
\usepackage{comment}
\usepackage{color}
\usepackage{soul}

\begin{document}

\author{Ido Almog$^1$}
\author{Gil Loewenthal$^1$}
\author{Jonathan Coslovsky$^1$}
\author{Yoav Sagi$^2$}
\author{Nir Davidson$^1$}

\address{$^1$Department of Physics of Complex Systems, Weizmann Institute of Science, Rehovot 76100, Israel}
\address{$^2$JILA, National Institute of Standards and Technology and University of Colorado, Department of Physics, University of Colorado, Boulder, Colorado 80309-0440, USA}

\title[]{Optimal dynamical decoupling in the presence of colored control noise}
\pacs{03.65.-w,03.65.Yz,03.67.-a,82.56.Jn}

\begin{abstract}
An optimal dynamical decoupling of a quantum system coupled to a noisy environment must take into account also the imperfections of the control pulses.
We present a new formalism which describes, in a closed-form expression, the evolution of the system, including the spectral function of both the environment and control noise. We show that by measuring these spectral functions, our expression can be used to optimize the decoupling pulse sequence. We demonstrate this approach with an ensemble of optically trapped ultra-cold Rubidium atoms, and use quantum process tomography to identify the effect of the environment and control noise. Our approach is applicable and important for any realistic implementation of quantum information processing.
\end{abstract}

\maketitle

\emph{Introduction - }The paradigm of a two-level system (TLS) is central to quantum information (QI), where it is applied as a quantum bit (qubit), the building block for information transfer, quantum memory or the computational gates.
A Coupling to a noisy environment, which is inherent to any system, reduces the purity of the TLS, and thus limits its usefulness for any QI application.
Several techniques have been developed to increase the quality of the quantum operation of the TLS, which is usually quantified by a measure called fidelity \cite{QCQI}.

One of these techniques is dynamical decoupling (DD), where a pulsed control field is used to couple the two levels of the TLS, and thus reduce their coupling to the environment \cite{Levitt2008}.
In QI, DD is used mainly to reduce the decay of the fidelity of a TLS, making it useful for longer times, as demonstrated in  a large variety of systems \cite{kotler_2011,our_process_tomography,Viola1999,PhysRevLett.82.2417,PhysRevLett.85.2272,PhysRevLett.95.180501,uhrig2007,Bylander2011}.
It was theoretically and experimentally shown that the success of these schemes can be predicted using a measurable spectral function that describes the coupling of the system to the environment, sometimes referred to as the ``bath spectrum'' \cite{Bylander2011,Kofman2001,gordon2007,Uys2009,direct,alvarez2011,biercuk2012,PhysRevLett.104.040401}.
In all these proposals, the higher the rate of the control pulses, the better is the decoupling from the environment.
However, in any realistic implementation there are imperfections in the control field, hence dynamical decoupling sequences become increasingly ineffective as the number of pulses grow.

In the quest for robust DD in the presence of  pulse imperfections, it is usually assumed that errors in the control pulses are uncorrelated \cite{PhysRevLett.90.037901,Ager2010,Souza2011}.
Such a ``white'' noise assumption is commonly used when estimating the fidelity of a gate, using the same gate many times but in a random manner, an approach known as ``benchmarking'' \cite{PhysRevLett.109.080505}.
In contrast to this assumption, as we show below, the control pulses often have correlated errors leading to a non-flat spectral function, which must be taken into account.

In this work, we study the combined effect of coupling to the environment and DD control field with colored noise on the TLS, and develop for it a closed-form expression, using the two corresponding spectral functions.
We measure these spectral functions with an ensemble of ultra-cold optically trapped $^{87}Rb$ atoms, and then use them to predict the outcome of a generic DD scheme and its overall fidelity.
Using quantum process tomography \cite{QCQI,our_process_tomography}, we show and explain the effect of each of the spectral functions with any initial state.

\emph{System subjected to realistic dynamical decoupling - } We consider a general TLS model with energy fluctuations and a noisy control field, described by the effective Hamiltonian:
\begin{equation}\label{Hamiltonian3}
\hat{H}=\frac{\hbar}{2}\left[\omega_0+\delta(t)\right]\sigma_z +\frac{\hbar}{2}\left[\Omega(t)e^{-i\omega_0 t}\sigma_x+h.c. \right] \ \ .
\end{equation}
Here, $\omega_0$ is the transition frequency between the two states and $\Omega(t)e^{-i\omega_0 t}$ is a noisy (classical) external control field, which is used for the DD.
The operators $\sigma_i$ are the Pauli matrices, written in the two level basis denoted by $\ket{2}$ and $\ket{1}$. The noise in the control field enters through $\Omega(t)=\Omega_0(t)\left[1+n_c(t)\right]$, where $\Omega_0(t)$ is the desired, noiseless control.
The frequency detuning noise, $\delta(t)$, and the control noise, $n_c(t)$, are random functions of time, with zero average.
Our analysis can be readily extended for control fields having both multiplicative and additive noise, multi-axis pulses, and frequency and phase noise in the control \cite{com_sup_noise}.

The short time evolution of an initial state can be described by the reduced density matrix: $\tilde{\rho}=\frac{1}{2}\left(\rho_x\sigma_x+\rho_y\sigma_y+\rho_z\sigma_z+\mathbf{I}\right)$, to second order in the noises \cite{com_sup_noise}.
In the interaction picture of $H_0(t)=\hbar\sigma_x\Omega_0(t)/2+\hbar\sigma_z\omega_0/2$, and under  the weak coupling assumption \cite{direct}, the effective Hamiltonian, $\tilde{H}_{int}(t)=\hbar\delta(t)\frac{\sigma_z}{2} cos(\int_0^t d\tau\Omega_0(\tau)/2)+\hbar n_c(t)\Omega_0(t)\frac{\sigma_x}{2}$, can be considered as a perturbation. The master equation for the density matrix operator is:

\begin{equation}
\label{gen-ME}
\dot{\tilde\rho}(t)=\expect{\frac{1}{(i \hbar)^2}\int_0^t dt_2
\left[\tilde{H}_{int}(t),\tilde{H}_{int}(t_2)\tilde{\rho}(t)\right]  + h.c. } \ \ ,
\end{equation}
where $\expect{...}$ stands for expectation value, after tracing out the environment.
The two-term interaction Hamiltonian plugged into Eq. \ref{gen-ME} gives rise to four terms, which are integrated over time $T$ to find the short-time evolution of the density matrix. The outcome is the paper's theoretical main result:

\begin{widetext}
\begin{eqnarray}\label{decay_rate_density_matrix}
\Delta\rho(T)&=&\int_{-\infty}^{\infty}{df\left[ - \underbrace{\frac{\rho_x\sigma_x+\rho_y\sigma_y}{4} G_{\delta}(f)F_{\delta}(f)}_{\text{coupling to the environment}}
 -\underbrace{\frac{\rho_y\sigma_y+\rho_z\sigma_z}{4}  G_{c}(f)F_{c}(f)}_{\text{noise in the control}}
  +\underbrace{ \left(\frac{\rho_z\sigma_x}{4} G_{\delta c}(f)  +  \frac{\rho_x\sigma_z}{4}G_{c \delta}(f) \right)F_{\delta c}(f)}_{\text{cross-correlation between environment and control}} \right]}\ \nonumber \\
\end {eqnarray}
\end{widetext}

The three spectral overlap integrals in Eq. \ref{decay_rate_density_matrix}, determine the full evolution of the density matrix. They describe (in this order) the effect of the coupling to the environment, the noise in the control field and cross-correlation between the environment and the control field.
The first term is similar to the spectral overlap integral of Refs. \cite{Kofman2001,PhysRevLett.104.040401,direct}.

The two bath spectral functions $G_{\delta}(f)$ and $G_{c}(f)$ (describe the correlation at different times of the environment and the noise of the control:
\begin{eqnarray}
\label{bath_function}
G_{\delta}(f)& \equiv &\int_{-\infty}^{\infty}{e^{-2\pi i f \tau}\expect{\delta(t)\cdot \delta(t+\tau)}d\tau} \nonumber \\
G_{c}(f)& \equiv &\int_{-\infty}^{\infty}{e^{-2\pi i f \tau}\expect{n_c(t)\cdot n_c(t+\tau)}d\tau} \ \ ,
\end{eqnarray}
The filter spectral functions $F_{\delta}(f)$ and $F_{c}(f)$ encapsulate the information regarding the modulation done by the control, during the time period $T$, and are written explicitly for sequences composed of $\pi_x$ or $\pi_{-x}$ pulses as:
\begin{eqnarray}
\label{eq_for_F_t}
F_{\delta}(f)& \equiv &\left\vert \int_0^T{dt e^{-2\pi i f t}\cdot \cos\left(\int_0^t{\Omega_0(\tau)d\tau}\right) } \right\vert ^2 \nonumber \\
F_c(f)& \equiv &\left\vert \int_0^T{dt e^{-2\pi i f t}\cdot \Omega_0(t)}\right\vert^2 \ \ .
\end{eqnarray}
Similarly, $F_{\delta c}(f)$, $G_{\delta c}(f)$ and  $G_{c \delta}(f)$, describing the cross-correlations between the control noise, $n_c(t)$, and the  environment noise, $\delta(t)$ are given in \cite{com_sup_noise}, but are negligible in our experiment.

Inverting the relation in Eq. \ref{decay_rate_density_matrix}, in order to find the bath spectral functions from time evolution measurements of the density matrix is hard, when two or more overlap integrals are involved.
However, it is possible to reduce the expressions to a single overlap integral by choosing wisely the initial state and DD sequence, essentially separating the problems of finding the two spectral functions.
For example, to find the environment bath spectrum, $G_{\delta}(f)$, we have used a random initial state and employed envelope spectroscopy to be insensitive to the control noise.
With this choice, the evolution of the reduced density matrix, as given in Eq. 3, depends only on a single overlap integral \cite{direct}.
Using a filter function consisting of several discrete peaks, which samples the environment bath spectrum at these discrete frequencies, one can invert the spectral overlap integral solving a linear set of equations \cite{Bylander2011} or a single equation in the case of a single peak filter function \cite{direct}.

Similarly, in order to measure the control noise spectral function, $G_{c}(f)$, it is worthy to cancel the overlap integral of the environment.
This is done by applying $\pi$-pulse sequence starting with an ensemble initialized to the state $\rho(0)=\ket{1}$.
Since it  essentially keeps the system in states  $\ket{1}$ and $\ket{2}$ that are insensitive to the pure dephasing environment noise, it reduces Eq. \ref{decay_rate_density_matrix} to \cite{com_sup_noise}: $\Delta\rho(T)=\frac{1}{4}\expect{\rho_y^2(T)} \sigma_z$ ,
with
\begin{equation}
\label{overlap_noise}
\expect{\rho_y^2(T)}=\int_{-\infty}^{\infty}df G_c(f)F_c(f) \ \ .
\end{equation}
By applying $\pi/2$-pulse followed by state detection, we can measure $\expect{\rho_y^2(T)}$, which is sensitive to the overlap integral, as explained.

\emph{Measurement of the control noise spectrum - } Our experimental setup is described elsewhere \cite{PhysRevLett.105.093001}. In short, $\sim 2.5\cdot 10^5$ ultra-cold $^{87}Rb$ atoms are confined by an external optical potential created by two $1.06\:\mu$m crossed laser beams.
The ensemble temperature is $1.7\:\mu$K, and it has a central density of $2\cdot 10^{13}\:$cm$^{-3}$.
The two metastable states $\ket{1}=\ket{F=1,m=-1}$ and $\ket{2}=\ket{F=2,m=+1}$ of the $5^2S_{1/2}$ manifold are chosen as the TLS.
The energy difference between these states is, to the first order, magnetic insensitive, at the applied magnetic field of $3.2\:$G \cite{harber2002}.
The control field is implemented using a two-photon MW-RF transition detuned  by $\Delta\sim 110\:$kHz from the $\ket{F=2,m=0}$ level, taking into account all energy shifts (differential AC Stark shift, second order magnetic shifts, mean-field interaction and MW dressing).
We measure the state of the atoms using fluorescence detection scheme \cite{our_process_tomography}.

The main sources for  the noises $\delta(t)$ and $n_c(t)$ are  well understood in our system.
The environment noise, $\delta(t)$, is due to the differential AC Stark shift of elastically colliding atoms in the optical dipole trap \cite{our_process_tomography}.
For each of the atoms, the environment is the atomic ensemble itself, randomizing the atomic trajectory following every elastic collision.
The noise in the control is due to magnetic fluctuations.
The magnetic noise enters through the single-photon detuning of the two-photon transition, $\Delta(t)$, which is first order magnetic field sensitive.
This fluctuating detuning changes the effective Rabi frequency of the two-photon transition,
\begin{equation}
\label{eq_2photon}
\Omega(t)=\Omega_1\Omega_2/2\Delta(t) \ \ ,
\end{equation}
where $\Omega_1$ and $\Omega_2$, the single photon Rabi frequency of the MW and RF fields, are essentially noiseless in our system.
Note, that since the two states of the TLS are magnetically insensitive, noise in the magnetic field affect only the control field, hence the cross-correlation term in Eq. \ref{decay_rate_density_matrix}, is negligible.
We expect to find a dominant contribution to the noise in $50\:$Hz and higher harmonics, arising from the electrical grid.

\begin{figure}
    \centerline{\includegraphics[width=8.6cm]{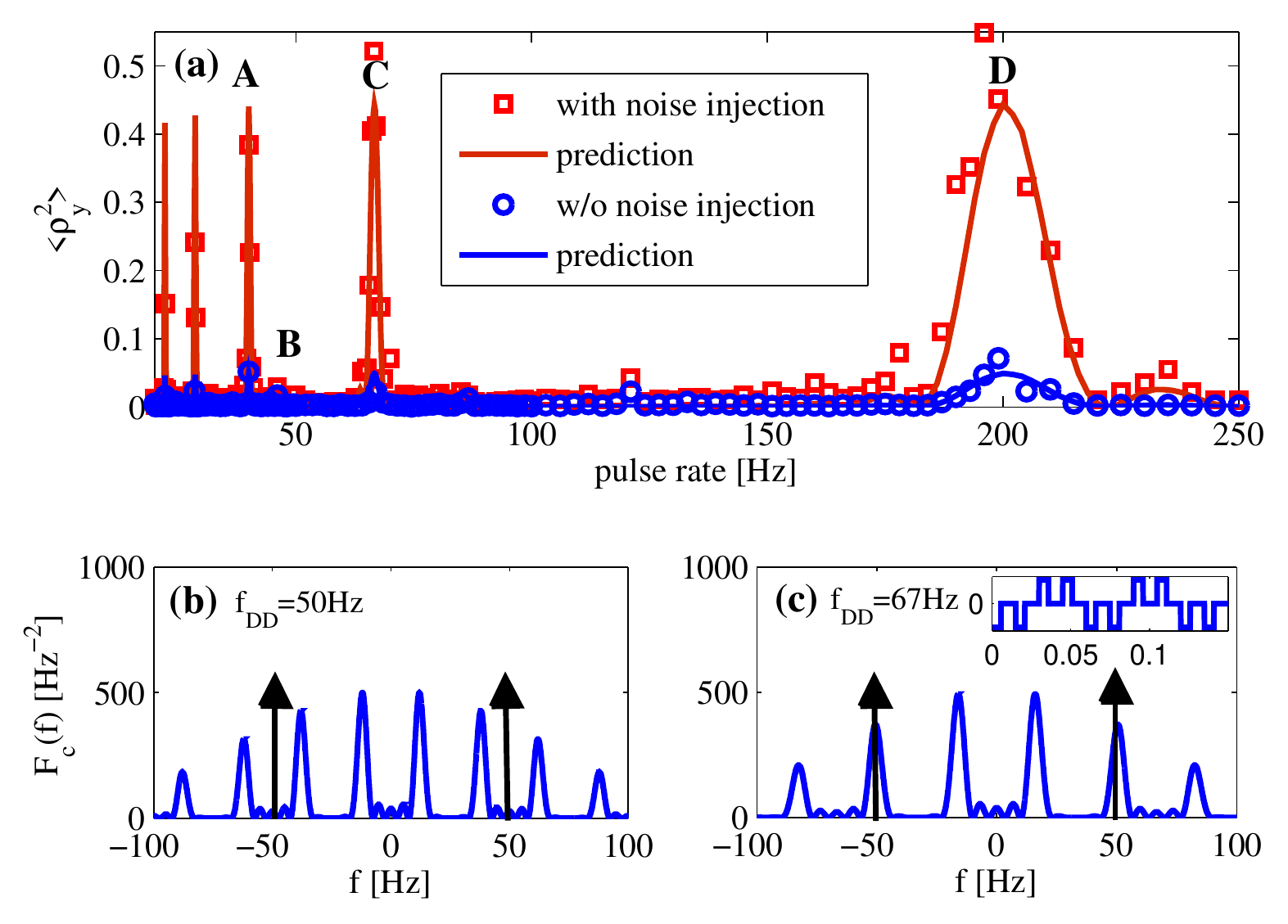}}
    \caption{The control noise spectrum.
    \textbf{(a)} The square (circle) data set is the measured $\expect{\rho_y^2}$ after 40 pulses of CPMG-4, with (without) injection of $4\:$mG magnetic noise at $50\:$Hz. The solid lines are calculated using Eqs. \ref{overlap_noise} and \ref{eq_2photon}, using the independently measured magnetic noise.
    Panels \textbf{(b)} and \textbf{(c)} : The filter function $F_c(f)$ of a CPMG-4 sequence with pulse rate of $50\:$Hz and $67\:$Hz, respectively. The arrows represent the $50\:$Hz component of the noise, which have significant  overlap with the peaks of the filter function only for the  sequence with a $67\:$Hz pulse rate (panel \textbf{(c)}).
The inset shows an illustration of a CPMG-4 control field at a $67\:$Hz pulse rate. The negative pulses are produced by a $\pi$ phase shift of the control field. (the duration of the pulses is not to scale). Points A, B, C and D are defined in the text.
    }
    \label{figure1}
\end{figure}

In Fig. ~\ref{figure1} \textbf {(a)} we plot the measured variance, $\expect{\rho_y^2}$, versus the pulse rate for 40 pulses of CPMG-4 DD sequence.
The sequence CPMG-n, initially introduced by Carr, Purcell, Meiboom, and Gill \cite{carr1954,meiboom:688}, is composed of equally spaced $\pi$ pulses with a phase alternating between $\pi$ and $-\pi$ after every $n/2$ pulses.
We repeat the measurements with and without deliberately injecting a $50\:$Hz magnetic noise (by driving a current in a nearby coil, phase locked to the electrical grid) to farther increase the control noise.
The measured spectrum is consistent with a single component at $50\:$Hz.
Notice that since the filter function of CPMG-4 at pulse rate of $50\:$Hz has no peak at this frequency, there is no special feature around $50\:$Hz, as illustrated in Fig. ~\ref{figure1} \textbf {(b)}.
This is in contrast to a pulse rate of $67\:$Hz which is overlapping a peak of the control noise, as shown in Fig. ~\ref{figure1} \textbf{(c)}.
Using Eqs. \ref{eq_for_F_t}-\ref{eq_2photon} and a direct independent measurement of the $50\:$Hz magnetic noise, we depict the calculated control noise spectrum (in ~\ref{figure1} \textbf {(a)}), showing discrete peaks at frequencies of $200/(2m-1)\:$Hz, $m=1,2,...$ in excellent agreement with the measured noise spectrum without any fit parameter.
The entire spectrum has a small bias which is due to imperfections in the state detection and uncorrelated (white) noise in the control pulses.
The latter is measured separately using a higher number of pulses (100), to increase the sensitivity.
Finally, a DC control  noise of $3\cdot 10^{-5}$, is measured using a CPMG DD sequence (with no phase alternation), whose filter function has a prominent component at DC.

By choosing one of the peaks in the spectrum of \ref{figure1} \textbf {(a)} and repeatedly measuring its noise component, we were able to reduce it by about 50 percent by injecting $50\:$Hz component to the nearby coil and searching for the phase and amplitude which minimize the peak.

\emph{DD Sequence engineering - } The usefulness of Eq. \ref{decay_rate_density_matrix} stems from its ability to predict the performance of any DD sequences, given that the two spectral functions, $G_\delta(f)$ and $G_c(f)$, are known.
For the purpose of optimizing the DD we quantify its success with the fidelity, defined by $\mathcal{F}=Tr\{\rho(t)\rho(0)\}$, as it includes both the effects of pulse imperfections and the coupling to the environment.
The short time fidelity (neglecting the last term in Eq. \ref{decay_rate_density_matrix}), can be written as \cite{com_sup_noise}:
\begin{eqnarray}
\label{decay_rate_fidelity}
\mathcal{F}&=&1 - \frac{\rho_x^2+\rho_y^2}{4} \int_{-\infty}^{\infty}dfG_{\delta}(f)F_{\delta}(f)  \nonumber \\
& &- \frac{\rho_y^2+\rho_z^2}{4} \int_{-\infty}^{\infty}df G_{c}(f)F_{c}(f)  \ \ .
\end{eqnarray}

\begin{figure}
    \centerline{\includegraphics[width=8.6cm]{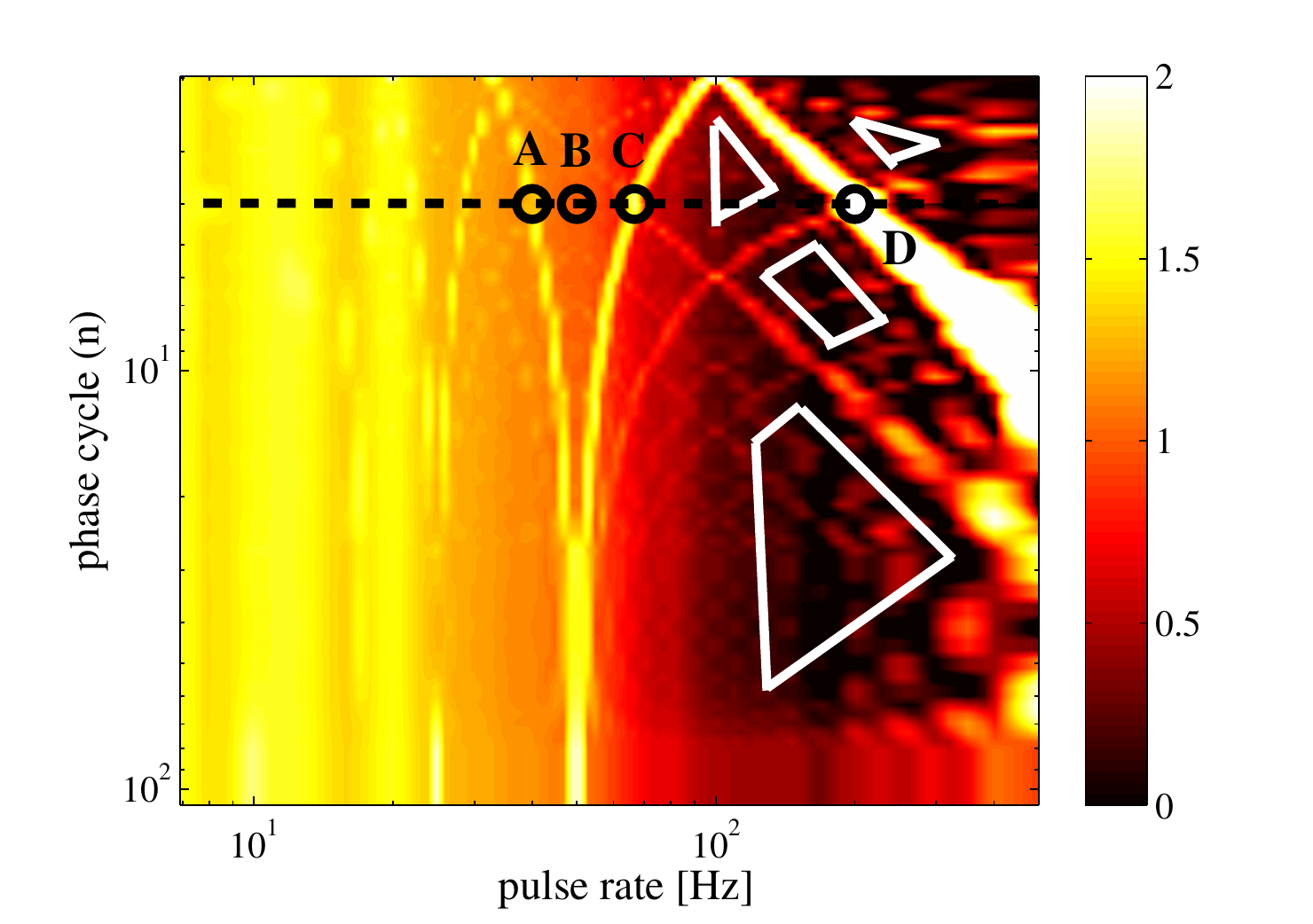}}
    \caption{
    A two-dimensional map of the predicted logarithm of fidelity decay rate (in units of ${\text{sec}}^{-1}$) presented in the variables pulse rate, $f_{DD}$ and phase cycle n of the CPMG-n pulse sequence.
    The shade (color online) represent the decay rate of the fidelity in logarithmic scale (black represents no decay). The marked zones of best expected performance of the DD sequence are identified by eye inspection. Points A, B, C and D are defined in the text. The dashed line correspond to $n=4$ (i.e. CPMG-4 sequence).
    }
    \label{map_2d}
\end{figure}

It is helpful to plot the decay rate of the fidelity as a function of the pulse sequence parameters.
In such a plot, it is easy to graphically identify the region in parameter space where the performance of the DD sequence is optimized.
In the case of CPMG-n, the natural choice of parametrization is the  pulse rate, $f_{DD}$ and the parameter n.
An example for such a calculated map, based on the two measured spectral bath functions in our system is presented in Fig. \ref{map_2d}, for the worst case fidelity (taking an initial state $\rho_y=1$).
The regions with the highest fidelity are clearly visible.

Although the map exhibits some complex features, the central ones can be qualitatively understood.
At low control pulse rate, the fidelity decay rate follows a Lorentzian, reflecting the Poisson statistics of the cold atomic collisions \cite{our_process_tomography}.
For higher pulse frequencies, there is a reduction in fidelity owing to white noise in the control field arising from the large number of pulses.
Large-$n$ cycles are also less successful, since their filter function has a large spectral component in DC, sampling the slow drifts in strength of the control field.
The control noise, originating mainly from the $50\:$Hz magnetic noise, produces a dominant feature appearing as strips on the map.
The points, A-D correspond to the points marked in Fig. ~\ref{figure1}.

\emph{Coherence of an arbitrary initial state - } Process tomography \cite{QCQI} is a technique used to characterize the TLS state after being manipulated by the control (referred here by {\em process}), for any initial state.
A great advantage of the formalism presented in Eq. \ref{decay_rate_density_matrix} is that it predicts the entire 3-dimensional effect of the process on the system, which is simply visualized as deformed sphere in the Bloch representation.

For process tomography we repeat the process with four initial states, $\rho(0)=\frac{1}{2}\left(\Sigma+\mathbf{1}\right)$, where $\Sigma$ is the Pauli matrix $\sigma_z$, $-\sigma_z$, $\sigma_x$ or $\sigma_y$.
For each initial state, the final state is measured by applying 6 different control pulses, followed by a state detection.
For a linear process, this information is sufficient to construct the process matrix \cite{QCQI,our_process_tomography}.

\begin{figure}
    \centerline{\includegraphics[width=8cm]{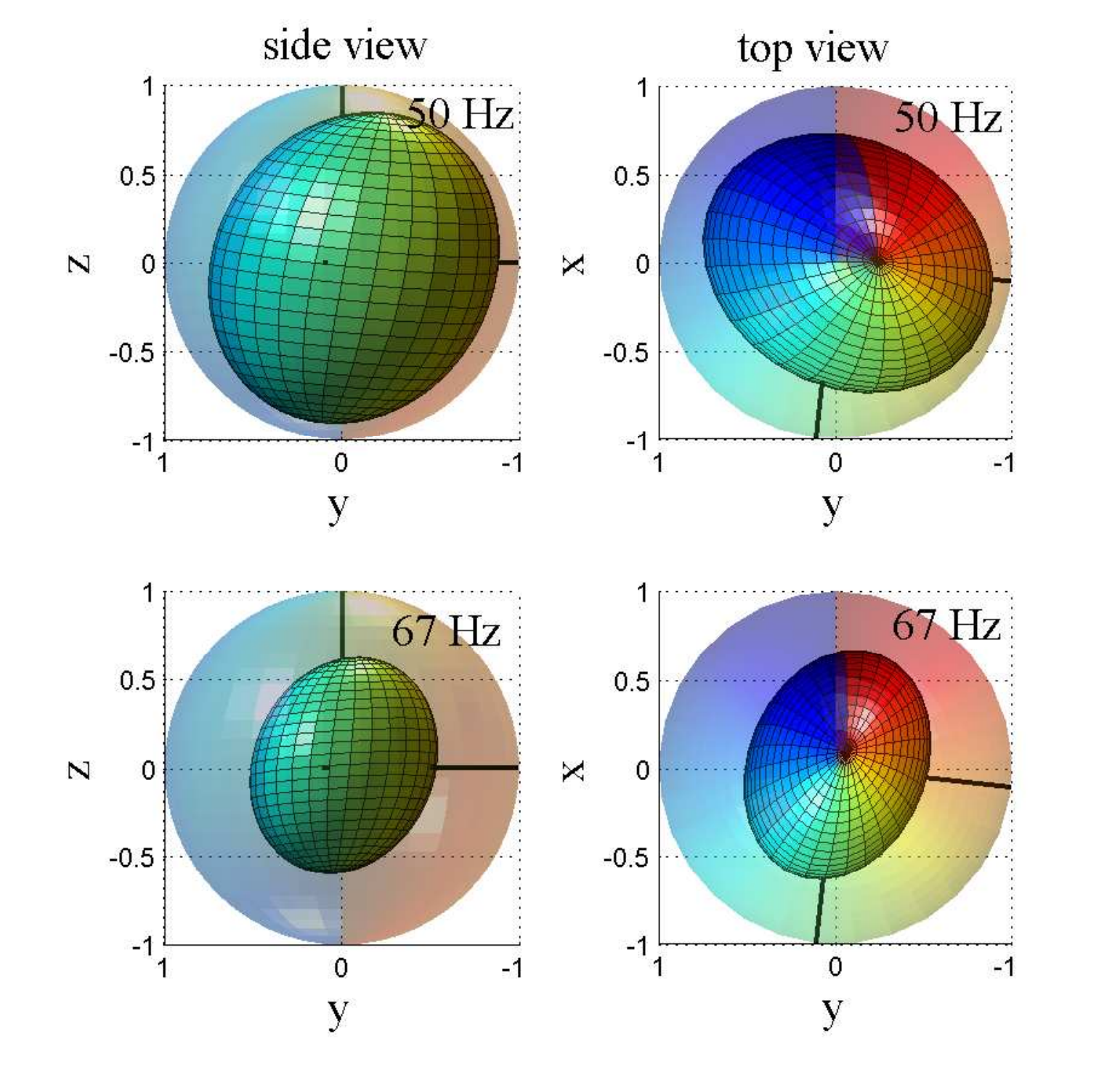}}
    \caption{Process tomography of CPMG-4 DD scheme in the presence of a noisy control field. A Color code is used to help determine the outcome with any initial state.
    Top: Measured Bloch sphere after 50 pulses at a rate of $50\:$Hz with 4 mG magnetic noise injection. Top and side view on the right and left respectively.
    Bottom: The same with $67\:$Hz DD.
    The different axes suffer from different decay rates in each case as expected from Eq. \ref{decay_rate_density_matrix}.
    }
    \label{figure_eggs}
\end{figure}

The results of the process tomography measurements are shown in Fig. \ref{figure_eggs}, for two DD processes: CPMG-4 at $50\:$Hz and $67\:$Hz pulse rates, corresponding to points B and C in Fig. \ref{map_2d}.
The two processes, although close in frequency, differ significantly, in agreement with our measured bath spectral functions.
For the $50\:$Hz process, there is no dominant control noise (as explained before), hence the decay is mostly due to the coupling to the environment. The decay rate of the z axis is the slowest, limited by a T1 process (not included in our model, measured to be $\sim 5$sec). The decay of the other two axes is similar, which is expected from Eq. \ref{decay_rate_density_matrix}, since the coefficients $\rho_x$ and $\rho_y$ appear symmetrically in the terms describing the coupling to the environment. In contrast, in the $67\:$Hz process, the decay in the y and z axes is faster since it is also affected by the noise in the control (involving the coefficients $\rho_y$ and $\rho_y$).

\emph{Conclusions - }
The formalism developed here together with the one described in \cite{direct} gives a recipe for designing a DD sequence:
First measure the spectral function defining the coupling to environment.
Then measure the spectral function that characterizes the noise in the control field.
Choose a general DD sequence parameterized by few parameters.
Use the overlap integrals to calculate the performance map as a function of these parameters.
Choose high fidelity regions for the DD sequences.
This framework can be extended also to sequences combining $\pi_x$ and $\pi_y$ pulses.
The filter functions, however, don't have an analytical expression, in this case.
Although the  the optimal pulse is system dependent, The most efficient way to find it is general, as we have shown here.

We acknowledge the financial support of MINERVA, ISF and DIP.

\bibliographystyle{apsrev4-1}

\begin{thebibliography}{27}%
\makeatletter
\providecommand \@ifxundefined [1]{%
 \@ifx{#1\undefined}
}%
\providecommand \@ifnum [1]{%
 \ifnum #1\expandafter \@firstoftwo
 \else \expandafter \@secondoftwo
 \fi
}%
\providecommand \@ifx [1]{%
 \ifx #1\expandafter \@firstoftwo
 \else \expandafter \@secondoftwo
 \fi
}%
\providecommand \natexlab [1]{#1}%
\providecommand \enquote  [1]{``#1''}%
\providecommand \bibnamefont  [1]{#1}%
\providecommand \bibfnamefont [1]{#1}%
\providecommand \citenamefont [1]{#1}%
\providecommand \href@noop [0]{\@secondoftwo}%
\providecommand \href [0]{\begingroup \@sanitize@url \@href}%
\providecommand \@href[1]{\@@startlink{#1}\@@href}%
\providecommand \@@href[1]{\endgroup#1\@@endlink}%
\providecommand \@sanitize@url [0]{\catcode `\\12\catcode `\$12\catcode
  `\&12\catcode `\#12\catcode `\^12\catcode `\_12\catcode `\%12\relax}%
\providecommand \@@startlink[1]{}%
\providecommand \@@endlink[0]{}%
\providecommand \url  [0]{\begingroup\@sanitize@url \@url }%
\providecommand \@url [1]{\endgroup\@href {#1}{\urlprefix }}%
\providecommand \urlprefix  [0]{URL }%
\providecommand \Eprint [0]{\href }%
\providecommand \doibase [0]{http://dx.doi.org/}%
\providecommand \selectlanguage [0]{\@gobble}%
\providecommand \bibinfo  [0]{\@secondoftwo}%
\providecommand \bibfield  [0]{\@secondoftwo}%
\providecommand \translation [1]{[#1]}%
\providecommand \BibitemOpen [0]{}%
\providecommand \bibitemStop [0]{}%
\providecommand \bibitemNoStop [0]{.\EOS\space}%
\providecommand \EOS [0]{\spacefactor3000\relax}%
\providecommand \BibitemShut  [1]{\csname bibitem#1\endcsname}%
\let\auto@bib@innerbib\@empty
%</preamble>
\bibitem [{\citenamefont {Nielsen}\ and\ \citenamefont {Chuang}(2000)}]{QCQI}%
  \BibitemOpen
  \bibfield  {author} {\bibinfo {author} {\bibfnamefont {M.~A.}\ \bibnamefont
  {Nielsen}}\ and\ \bibinfo {author} {\bibfnamefont {I.~L.}\ \bibnamefont
  {Chuang}},\ }\href@noop {} {\emph {\bibinfo {title} {Quantum Computation and
  Quantum Information}}}\ (\bibinfo  {publisher} {Cambridge University Press},\
  \bibinfo {year} {2000})\BibitemShut {NoStop}%
\bibitem [{\citenamefont {Levitt}(2008)}]{Levitt2008}%
  \BibitemOpen
  \bibfield  {author} {\bibinfo {author} {\bibfnamefont {M.~H.}\ \bibnamefont
  {Levitt}},\ }\href@noop {} {\emph {\bibinfo {title} {Spin Dynamics: Basics of
  Nuclear Magnetic Resonance}}}\ (\bibinfo  {publisher} {Wiley-Blackwell},\
  \bibinfo {year} {2008})\BibitemShut {NoStop}%
\bibitem [{\citenamefont {Kotler}\ \emph {et~al.}(2011)\citenamefont {Kotler},
  \citenamefont {Akerman}, \citenamefont {Glickman}, \citenamefont {Keselman},\
  and\ \citenamefont {Ozeri}}]{kotler_2011}%
  \BibitemOpen
  \bibfield  {author} {\bibinfo {author} {\bibfnamefont {S.}~\bibnamefont
  {Kotler}}, \emph {et~al.},\ } {\bibfield  {journal} {\bibinfo
  {journal} {Nature}\ }\textbf {\bibinfo {volume} {473}},\ \bibinfo {pages}
  {61} (\bibinfo {year} {2011})}\BibitemShut {NoStop}%
\bibitem [{\citenamefont {Sagi}\ \emph
  {et~al.}(2010{\natexlab{a}})\citenamefont {Sagi}, \citenamefont {Almog},\
  and\ \citenamefont {Davidson}}]{our_process_tomography}%
  \BibitemOpen
  \bibfield  {author} {\bibinfo {author} {\bibfnamefont {Y.}~\bibnamefont
  {Sagi}}, \bibinfo {author} {\bibfnamefont {I.}~\bibnamefont {Almog}}, \ and\
  \bibinfo {author} {\bibfnamefont {N.}~\bibnamefont {Davidson}},\ } {\bibfield  {journal} {\bibinfo
  {journal} {Phys. Rev. Lett.}\ }\textbf {\bibinfo {volume} {105}},\ \bibinfo
  {pages} {053201} (\bibinfo {year} {2010}{\natexlab{a}})}\BibitemShut
  {NoStop}%
\bibitem [{\citenamefont {Viola}\ \emph
  {et~al.}(1999{\natexlab{a}})\citenamefont {Viola}, \citenamefont {Lloyd},\
  and\ \citenamefont {Knill}}]{Viola1999}%
  \BibitemOpen
  \bibfield  {author} {\bibinfo {author} {\bibfnamefont {L.}~\bibnamefont
  {Viola}}, \bibinfo {author} {\bibfnamefont {S.}~\bibnamefont {Lloyd}}, \ and\
  \bibinfo {author} {\bibfnamefont {E.}~\bibnamefont {Knill}},\ }\href@noop {}
  {\bibfield  {journal} {\bibinfo  {journal} {Phys. Rev. Lett.}\ }\textbf
  {\bibinfo {volume} {83}},\ \bibinfo {pages} {4888} (\bibinfo {year}
  {1999}{\natexlab{a}})}\BibitemShut {NoStop}%
\bibitem [{\citenamefont {Viola}\ \emph
  {et~al.}(1999{\natexlab{b}})\citenamefont {Viola}, \citenamefont {Knill},\
  and\ \citenamefont {Lloyd}}]{PhysRevLett.82.2417}%
  \BibitemOpen
  \bibfield  {author} {\bibinfo {author} {\bibfnamefont {L.}~\bibnamefont
  {Viola}}, \bibinfo {author} {\bibfnamefont {E.}~\bibnamefont {Knill}}, \ and\
  \bibinfo {author} {\bibfnamefont {S.}~\bibnamefont {Lloyd}},\ }{\bibfield  {journal} {\bibinfo
  {journal} {Phys. Rev. Lett.}\ }\textbf {\bibinfo {volume} {82}},\ \bibinfo
  {pages} {2417} (\bibinfo {year} {1999}{\natexlab{b}})}\BibitemShut {NoStop}%
\bibitem [{\citenamefont {Search}\ and\ \citenamefont
  {Berman}(2000)}]{PhysRevLett.85.2272}%
  \BibitemOpen
  \bibfield  {author} {\bibinfo {author} {\bibfnamefont {C.}~\bibnamefont
  {Search}}\ and\ \bibinfo {author} {\bibfnamefont {P.~R.}\ \bibnamefont
  {Berman}},\ } {\bibfield
  {journal} {\bibinfo  {journal} {Phys. Rev. Lett.}\ }\textbf {\bibinfo
  {volume} {85}},\ \bibinfo {pages} {2272} (\bibinfo {year}
  {2000})}\BibitemShut {NoStop}%
\bibitem [{\citenamefont {Khodjasteh}\ and\ \citenamefont
  {Lidar}(2005)}]{PhysRevLett.95.180501}%
  \BibitemOpen
  \bibfield  {author} {\bibinfo {author} {\bibfnamefont {K.}~\bibnamefont
  {Khodjasteh}}\ and\ \bibinfo {author} {\bibfnamefont {D.~A.}\ \bibnamefont
  {Lidar}},\ } {\bibfield
  {journal} {\bibinfo  {journal} {Phys. Rev. Lett.}\ }\textbf {\bibinfo
  {volume} {95}},\ \bibinfo {pages} {180501} (\bibinfo {year}
  {2005})}\BibitemShut {NoStop}%
\bibitem [{\citenamefont {Uhrig}(2007)}]{uhrig2007}%
  \BibitemOpen
  \bibfield  {author} {\bibinfo {author} {\bibfnamefont {G.~S.}\ \bibnamefont
  {Uhrig}},\ }{\bibfield
  {journal} {\bibinfo  {journal} {Physical Review Letters}\ }\textbf {\bibinfo
  {volume} {98}},\ \bibinfo {eid} {100504} (\bibinfo {year}
  {2007})}\BibitemShut {NoStop}%
\bibitem [{\citenamefont {Bylander}\ \emph {et~al.}(2011)\citenamefont
  {Bylander}, \citenamefont {Gustavsson}, \citenamefont {Yan}, \citenamefont
  {Yoshihara}, \citenamefont {Harrabi}, \citenamefont {Fitch}, \citenamefont
  {Cory}, \citenamefont {Nakamura}, \citenamefont {Tsai},\ and\ \citenamefont
  {Oliver}}]{Bylander2011}%
  \BibitemOpen
  \bibfield  {author} {\bibinfo {author} {\bibfnamefont {J.}~\bibnamefont
  {Bylander}}, \emph {et~al.},\ }
  {\bibfield  {journal} {\bibinfo  {journal} {Nat Phys}\ }\textbf {\bibinfo
  {volume} {7}},\ \bibinfo {pages} {565} (\bibinfo {year} {2011})}\BibitemShut
  {NoStop}%
\bibitem [{\citenamefont {Kofman}\ and\ \citenamefont
  {Kurizki}(2001)}]{Kofman2001}%
  \BibitemOpen
  \bibfield  {author} {\bibinfo {author} {\bibfnamefont {A.~G.}\ \bibnamefont
  {Kofman}}\ and\ \bibinfo {author} {\bibfnamefont {G.}~\bibnamefont
  {Kurizki}},\ } {\bibfield
  {journal} {\bibinfo  {journal} {Phys. Rev. Lett.}\ }\textbf {\bibinfo
  {volume} {87}},\ \bibinfo {pages} {270405} (\bibinfo {year}
  {2001})}\BibitemShut {NoStop}%
\bibitem [{\citenamefont {Gordon}\ \emph {et~al.}(2007)\citenamefont {Gordon},
  \citenamefont {Erez},\ and\ \citenamefont {Kurizki}}]{gordon2007}%
  \BibitemOpen
  \bibfield  {author} {\bibinfo {author} {\bibfnamefont {G.}~\bibnamefont
  {Gordon}}, \bibinfo {author} {\bibfnamefont {N.}~\bibnamefont {Erez}}, \ and\
  \bibinfo {author} {\bibfnamefont {G.}~\bibnamefont {Kurizki}},\ } {\bibfield  {journal}
  {\bibinfo  {journal} {Journal of Physics B: Atomic, Molecular and Optical
  Physics}\ }\textbf {\bibinfo {volume} {40}},\ \bibinfo {pages} {S75}
  (\bibinfo {year} {2007})}\BibitemShut {NoStop}%
\bibitem [{\citenamefont {Uys}\ \emph {et~al.}(2009)\citenamefont {Uys},
  \citenamefont {Biercuk},\ and\ \citenamefont {Bollinger}}]{Uys2009}%
  \BibitemOpen
  \bibfield  {author} {\bibinfo {author} {\bibfnamefont {H.}~\bibnamefont
  {Uys}}, \bibinfo {author} {\bibfnamefont {M.~J.}\ \bibnamefont {Biercuk}}, \
  and\ \bibinfo {author} {\bibfnamefont {J.~J.}\ \bibnamefont {Bollinger}},\
  } {\bibfield  {journal}
  {\bibinfo  {journal} {Phys. Rev. Lett.}\ }\textbf {\bibinfo {volume} {103}},\
  \bibinfo {pages} {040501} (\bibinfo {year} {2009})}\BibitemShut {NoStop}%
\bibitem [{\citenamefont {Almog}\ \emph {et~al.}(2011)\citenamefont {Almog},
  \citenamefont {Sagi}, \citenamefont {Gordon}, \citenamefont {Bensky},
  \citenamefont {Kurizki},\ and\ \citenamefont {Davidson}}]{direct}%
  \BibitemOpen
  \bibfield  {author} {\bibinfo {author} {\bibfnamefont {I.}~\bibnamefont
  {Almog}}, \emph {et~al.},\ } {\bibfield  {journal}
  {\bibinfo  {journal} {Journal of Physics B: Atomic, Molecular and Optical
  Physics}\ }\textbf {\bibinfo {volume} {44}},\ \bibinfo {pages} {154006}
  (\bibinfo {year} {2011})}\BibitemShut {NoStop}%
\bibitem [{\citenamefont {Alvarez}\ and\ \citenamefont
  {Suter}(2011)}]{alvarez2011}%
  \BibitemOpen
  \bibfield  {author} {\bibinfo {author} {\bibfnamefont {G.~A.}\ \bibnamefont
  {Alvarez}}\ and\ \bibinfo {author} {\bibfnamefont {D.}~\bibnamefont
  {Suter}},\ } {\bibfield
  {journal} {\bibinfo  {journal} {Phys. Rev. Lett.}\ }\textbf {\bibinfo
  {volume} {107}},\ \bibinfo {pages} {230501} (\bibinfo {year}
  {2011})}\BibitemShut {NoStop}%
\bibitem [{\citenamefont {Green}\ \emph {et~al.}(2012)\citenamefont {Green},
  \citenamefont {Uys},\ and\ \citenamefont {Biercuk}}]{biercuk2012}%
  \BibitemOpen
  \bibfield  {author} {\bibinfo {author} {\bibfnamefont {T.}~\bibnamefont
  {Green}}, \bibinfo {author} {\bibfnamefont {H.}~\bibnamefont {Uys}}, \ and\
  \bibinfo {author} {\bibfnamefont {M.~J.}\ \bibnamefont {Biercuk}},\ } {\bibfield  {journal} {\bibinfo
  {journal} {Phys. Rev. Lett.}\ }\textbf {\bibinfo {volume} {109}},\ \bibinfo
  {pages} {020501} (\bibinfo {year} {2012})}\BibitemShut {NoStop}%
\bibitem [{\citenamefont {Clausen}\ \emph {et~al.}(2010)\citenamefont
  {Clausen}, \citenamefont {Bensky},\ and\ \citenamefont
  {Kurizki}}]{PhysRevLett.104.040401}%
  \BibitemOpen
  \bibfield  {author} {\bibinfo {author} {\bibfnamefont {J.}~\bibnamefont
  {Clausen}}, \bibinfo {author} {\bibfnamefont {G.}~\bibnamefont {Bensky}}, \
  and\ \bibinfo {author} {\bibfnamefont {G.}~\bibnamefont {Kurizki}},\ } {\bibfield  {journal} {\bibinfo
  {journal} {Phys. Rev. Lett.}\ }\textbf {\bibinfo {volume} {104}},\ \bibinfo
  {pages} {040401} (\bibinfo {year} {2010})}\BibitemShut {NoStop}%
\bibitem [{\citenamefont {Viola}\ and\ \citenamefont
  {Knill}(2003)}]{PhysRevLett.90.037901}%
  \BibitemOpen
  \bibfield  {author} {\bibinfo {author} {\bibfnamefont {L.}~\bibnamefont
  {Viola}}\ and\ \bibinfo {author} {\bibfnamefont {E.}~\bibnamefont {Knill}},\
  } {\bibfield  {journal}
  {\bibinfo  {journal} {Phys. Rev. Lett.}\ }\textbf {\bibinfo {volume} {90}},\
  \bibinfo {pages} {037901} (\bibinfo {year} {2003})}\BibitemShut {NoStop}%
\bibitem [{\citenamefont {{Tyryshkin}}\ \emph {et~al.}(2010)\citenamefont
  {{Tyryshkin}}, \citenamefont {{Wang}}, \citenamefont {{Zhang}}, \citenamefont
  {{Haller}}, \citenamefont {{Ager}}, \citenamefont {{Dobrovitski}},\ and\
  \citenamefont {{Lyon}}}]{Ager2010}%
  \BibitemOpen
  \bibfield  {author} {\bibinfo {author} {\bibfnamefont {A.~M.}\ \bibnamefont
  {{Tyryshkin}}}, \emph {et~al.},\ }\href@noop {}
  {\bibfield  {journal} (\bibinfo
  {year} {2010})},\ {arXiv:1011.1903
  [quant-ph]} \BibitemShut {NoStop}%
\bibitem [{\citenamefont {Souza}\ \emph {et~al.}(2011)\citenamefont {Souza},
  \citenamefont {\'Alvarez},\ and\ \citenamefont {Suter}}]{Souza2011}%
  \BibitemOpen
  \bibfield  {author} {\bibinfo {author} {\bibfnamefont {A.~M.}\ \bibnamefont
  {Souza}}, \bibinfo {author} {\bibfnamefont {G.~A.}\ \bibnamefont
  {\'Alvarez}}, \ and\ \bibinfo {author} {\bibfnamefont {D.}~\bibnamefont
  {Suter}},\ } {\bibfield
  {journal} {\bibinfo  {journal} {Phys. Rev. Lett.}\ }\textbf {\bibinfo
  {volume} {106}},\ \bibinfo {pages} {240501} (\bibinfo {year}
  {2011})}\BibitemShut {NoStop}%
\bibitem [{\citenamefont {Magesan}\ \emph {et~al.}(2012)\citenamefont
  {Magesan}, \citenamefont {Gambetta}, \citenamefont {Johnson}, \citenamefont
  {Ryan}, \citenamefont {Chow}, \citenamefont {Merkel}, \citenamefont
  {da~Silva}, \citenamefont {Keefe}, \citenamefont {Rothwell}, \citenamefont
  {Ohki}, \citenamefont {Ketchen},\ and\ \citenamefont
  {Steffen}}]{PhysRevLett.109.080505}%
  \BibitemOpen
  \bibfield  {author} {\bibinfo {author} {\bibfnamefont {E.}~\bibnamefont
  {Magesan}}, \emph {et~al.},\ }{\bibfield  {journal} {\bibinfo  {journal}
  {Phys. Rev. Lett.}\ }\textbf {\bibinfo {volume} {109}},\ \bibinfo {pages}
  {080505} (\bibinfo {year} {2012})}\BibitemShut {NoStop}%
\bibitem [{com()}]{com_sup_noise}%
  \BibitemOpen
  \href@noop {} {\bibinfo {pages} {For more details see the supplementary material section}}\BibitemShut
  {NoStop}%
%\bibitem [{\citenamefont {Gordon}\ and\ \citenamefont
%  {Kurizki}(2007)}]{PhysRevA.76.042310}%
%  \BibitemOpen
%  \bibfield  {author} {\bibinfo {author} {\bibfnamefont {G.}~\bibnamefont
%  {Gordon}}\ and\ \bibinfo {author} {\bibfnamefont {G.}~\bibnamefont
%  {Kurizki}},\ }\href {\doibase 10.1103/PhysRevA.76.042310} {\bibfield
%  {journal} {\bibinfo  {journal} {Phys. Rev. A}\ }\textbf {\bibinfo {volume}
%  {76}},\ \bibinfo {pages} {042310} (\bibinfo {year} {2007})}\BibitemShut
%  {NoStop}%
\bibitem [{\citenamefont {Sagi}\ \emph
  {et~al.}(2010{\natexlab{b}})\citenamefont {Sagi}, \citenamefont {Almog},\
  and\ \citenamefont {Davidson}}]{PhysRevLett.105.093001}%
  \BibitemOpen
  \bibfield  {author} {\bibinfo {author} {\bibfnamefont {Y.}~\bibnamefont
  {Sagi}}, \bibinfo {author} {\bibfnamefont {I.}~\bibnamefont {Almog}}, \ and\
  \bibinfo {author} {\bibfnamefont {N.}~\bibnamefont {Davidson}},\ } {\bibfield  {journal} {\bibinfo
  {journal} {Phys. Rev. Lett.}\ }\textbf {\bibinfo {volume} {105}},\ \bibinfo
  {pages} {093001} (\bibinfo {year} {2010}{\natexlab{b}})}\BibitemShut
  {NoStop}%
\bibitem [{\citenamefont {Harber}\ \emph {et~al.}(2002)\citenamefont {Harber},
  \citenamefont {Lewandowski}, \citenamefont {McGuirk},\ and\ \citenamefont
  {Cornell}}]{harber2002}%
  \BibitemOpen
  \bibfield  {author} {\bibinfo {author} {\bibfnamefont {D.~M.}\ \bibnamefont
  {Harber}}, \emph {et~al.},\ }{\bibfield
  {journal} {\bibinfo  {journal} {Phys. Rev. A}\ }\textbf {\bibinfo {volume}
  {66}},\ \bibinfo {pages} {053616} (\bibinfo {year} {2002})}\BibitemShut
  {NoStop}%
\bibitem [{\citenamefont {Carr}\ and\ \citenamefont
  {Purcell}(1954)}]{carr1954}%
  \BibitemOpen
  \bibfield  {author} {\bibinfo {author} {\bibfnamefont {H.~Y.}\ \bibnamefont
  {Carr}}\ and\ \bibinfo {author} {\bibfnamefont {E.~M.}\ \bibnamefont
  {Purcell}},\ }{\bibfield  {journal}
  {\bibinfo  {journal} {Phys. Rev.}\ }\textbf {\bibinfo {volume} {94}},\
  \bibinfo {pages} {630} (\bibinfo {year} {1954})}\BibitemShut {NoStop}%
\bibitem [{\citenamefont {Meiboom}\ and\ \citenamefont
  {Gill}(1958)}]{meiboom:688}%
  \BibitemOpen
  \bibfield  {author} {\bibinfo {author} {\bibfnamefont {S.}~\bibnamefont
  {Meiboom}}\ and\ \bibinfo {author} {\bibfnamefont {D.}~\bibnamefont {Gill}},\
  } {\bibfield  {journal} {\bibinfo
  {journal} {Rev. Sci. Instrum.}\ }\textbf {\bibinfo {volume} {29}},\ \bibinfo
  {pages} {688} (\bibinfo {year} {1958})}\BibitemShut {NoStop}%
\end{thebibliography}

\clearpage

\begin{widetext}


\section*{Supplementary material}

# EQUATION OF MOTION FOR SYSTEM DENSITY OPERATOR

### Mathematical derivation

We are interested in the short time evolution of the system density operator determined by the Hamiltonian:

$$\hat{H} = \frac{\hbar}{2}\left[\omega_0 + \delta(t)\right]\sigma_z + \frac{\hbar}{2}\left[\Omega(t)e^{-i\omega_0 t}\sigma_x + h.c.\right] \quad . \tag{1}$$

It is convenient to change to the interaction picture of $H_0(t) = \hbar\sigma_x\Omega_0(t)/2 + \hbar\sigma_z\omega_0/2$. In this picture, when the change in the system density operator during the bath correlation time is negligible, the effective Hamiltonian, $\tilde{H}_{int}$, can be considered as a perturbation. This assumption, called the *weak coupling* assumption, is naturally valid in QI since the quantum gate or memory is not useful when the decay of coherence is not small comparing to unity. In this case, the system's density obeys the following master-equation (to second order in the noises):

$$\dot{\tilde{\rho}}(t) = \left\langle \frac{1}{(i\hbar)^2}\int_0^t dt_2 \left[\tilde{H}_{int}(t),\left[\tilde{H}_{int}(t_2),\tilde{\rho}(t)\right]\right]\right\rangle \quad , \tag{2}$$

where $\langle...\rangle$ stands for expectation value, coming out from tracing out the environment.

The interaction hamiltonian can be written explicitly by $\tilde{H}_{int}(t) = \hbar\delta(t)\tilde{S}_\delta(t) + \hbar n_c(t)\tilde{S}_c(t)$, with an operator responsible for the interaction with the environment:

$$\tilde{S}_\delta(t) = e^{i\sigma_x \int_0^t d\tau\Omega_0(\tau)/2}\frac{\sigma_z}{2}e^{-i\sigma_x \int_0^t d\tau\Omega_0(\tau)/2} = \frac{\sigma_z}{2}\cos(\vartheta(t)) + \frac{\sigma_y}{2}\sin(\vartheta(t)) \tag{3}$$

and an operator responsible for the noise in the control,

$$\tilde{S}_c(t) = e^{i\sigma_x \int_0^t d\tau\Omega_0(\tau)/2}\frac{\Omega_0(t)\sigma_x}{2}e^{-i\sigma_x \int_0^t d\tau\Omega_0(\tau)/2} = \Omega_0(t)\frac{\sigma_x}{2} \quad , \tag{4}$$

where the angle $\vartheta(t) = \int_0^t d\tau\Omega_0(\tau)/2$ is sometimes called the action. Plugging this into Eq. 2 we get the equation for the density operator:

$$\begin{aligned}\dot{\tilde{\rho}}(t) = &-\int_0^t dt_2 \langle\delta(t)\delta(t_2)\rangle\left[\tilde{S}_\delta(t),\left[\tilde{S}_\delta(t_2),\tilde{\rho}(t)\right]\right] - \int_0^t dt_2 \langle n_c(t)n_c(t_2)\rangle\left[\tilde{S}_c(t),\left[\tilde{S}_c(t_2),\tilde{\rho}(t)\right]\right]\\ &-\int_0^t dt_2 \langle n_c(t)\delta(t_2)\rangle\left[\tilde{S}_c(t),\left[\tilde{S}_\delta(t_2),\tilde{\rho}(t)\right]\right] - \int_0^t dt_2 \langle\delta(t)n_c(t_2)\rangle\left[\tilde{S}_\delta(t),\left[\tilde{S}_c(t_2),\tilde{\rho}(t)\right]\right] \quad .\end{aligned} \tag{5}$$

Three mechanisms responsible for the change of the density matrix can be identified from the four terms in the expression: The first term describes the effect of the environment, the second describes the effect of the noise in the control field and the last two are mixed effect resulting from correlation between the control noise and the environment coupling. Fortunately, each term in Eq. 5 can be expressed as an overlap between two function in the spectral domain. For example the last term can be written:

$$-\int_0^t dt_2 \langle\delta(t)n_c(t_2)\rangle\left[\tilde{S}_\delta(t),\left[\tilde{S}_c(t_2)\tilde{\rho}(t)\right]\right] = \int_0^t dt_2 \langle\delta(t)n_c(t_2)\rangle\frac{1}{2}\sigma_x\Omega_0(t_2)\left[\rho_y(0)\sin(\vartheta(t)) + \rho_z(0)\cos(\vartheta(t))\right] \quad .\tag{6}$$

We integrate to get the expression for density matrix change contributed by this term:

$$\Delta\rho_{\delta c}(T) = \int_0^T dt_1 \int_0^{t_1} dt_2 \Phi_{\delta c}(t_1 - t_2)\cdot\frac{1}{2}\sigma_x\Omega_0(t_2)\rho_z(0)\cos(\vartheta(t_1)) \quad . \tag{7}$$

Note that to get this we ignored the $\sin(...)$ term since we constrain ourselves to sequences consisting on $\pi_x$ pulses only. Here we used the definition for the correlator which is a function of the time difference only: $\Phi_{\delta c}(t_1 - t_2) = \langle\delta(t_1)n_c(t_2)\rangle\Theta(t_1 - t_2)$, with heaviside function $\Theta(t_1 - t_2)$, looking forward to the next step where we replace the upper integral limit. Identifying a convolution integral between the functions $\Phi_{\delta c}(t)$ and $\Omega_0(t)$, with the notation $\Xi(t) = \cos(\vartheta(t))\Theta(t)\Theta(T - t)$, we write:

$$\Delta\rho_{\delta c}(T) = \frac{1}{2}\sigma_x\rho_z(0)\int_0^T dt_1\Xi(t_1)\left(\Phi_{\delta c}(t_1) * \Omega_0(t_1)\right) \quad . \tag{8}$$

Using the Fourier transform of the correlator (defined with a 2 factor to compensate for that the transformation of only the positive times) $G_{\delta c}(f) = 2 \int_{-\infty}^{\infty} d\tau e^{-2\pi i f \tau} \Phi_{\delta c}(\tau)$ and the symbols ~ for Fourier of other functions, the integral (without the coefficient) becomes

$$\Delta \rho_{\delta c}(T) = \frac{1}{2} \int_0^T dt_1 \Xi(t_1) \Phi_{\delta c}(t_1) * \Omega_0(t_1) = \frac{1}{4} \int_{-\infty}^{\infty} dt_1 \Xi(t_1) \int_{-\infty}^{\infty} df_1 e^{2\pi i f_1 t_1} G_{\delta c}(f_1) \cdot \tilde{\Omega}_0(f_1)$$

$$= \frac{1}{4} \int_{-\infty}^{\infty} dt \int_{-\infty}^{\infty} df_2 e^{2\pi i f_2 t} \tilde{\Xi}(f_2) \int_{-\infty}^{\infty} df_1 e^{2\pi i f_1 t} G_{\delta c}(f_1) \tilde{\Omega}_0(f_1) = \frac{1}{4} \int_{-\infty}^{\infty} df \tilde{\Xi}^*(f) \tilde{\Omega}_0(f) G_{\delta c}(f) \quad . \quad (9)$$

With the notation $F_{\delta c}(f) = \tilde{\Xi}^*(f)\tilde{\Omega}_0(f)$, the density matrix change resulting from this contribution:

$$\Delta \rho_{\delta c}(T) = \frac{1}{4} \sigma_x \rho_z(0) \int_{-\infty}^{\infty} df G_{\delta c}(f) F_{\delta c}(f) \quad . \quad (10)$$

Repeating this procedure taking the three other terms in Eq. 5, obtain the other overlap terms. The first two terms, describing the uncorrelated effect of the coupling to the environment and the noise in the control, are written explicitly in the Letter. The forth term is similar to Eq. 10 and can be written as:

$$\Delta \rho_{c\delta}(T) = \frac{1}{4} \sigma_z \rho_x(0) \int_{-\infty}^{\infty} df G_{c\delta}(f) F_{\delta c}(f) \quad , \quad (11)$$

with $G_{c\delta}(f)$ defined the same as $G_{\delta c}(f)$ up to exchanging the function $\delta(t)$ and $n_c(t)$.

### Generalization for other cases

Frequency (and phase) noise of the control are treated the same as the energy fluctuation $\delta(t)$. To see this, one has to change to the interaction picture which rotates with the frequency of the control local oscillator, including its classical frequency noise. In this interaction picture, the noise in the control is an additional $\sigma_z$ noise (up to a minus sign) added to $\delta(t)$.

Control pulses involving the two axes x and y should be considered carefully. Although changing the axis is equivalent to a phase shift, it cannot be regarded as energy fluctuations like is done with the frequency noise of the control, since the phase shift cannot be considered small, breaking the weak coupling assumption. Instead, one should carry on the similar derivation derivation as was done above. Applying this procedure, the commutators in Eq. 5 become somewhat more complicated to calculate since there are no close expression for $\tilde{S}_\delta(t)$ and $\tilde{S}_c(t)$, since the pulse action now includes the two non-commuting operators, $\sigma_x$ and $\sigma_y$. Nevertheless, $\tilde{S}_\delta(t)$ and $\tilde{S}_c(t)$ can be calculated directly, by evaluating the effect the pulses, sequentially. Since between each pulse the evolution can be calculated analytically, the resulting filter function can be written as a as long table of function as the number of pulses, making it still possible to calculate the overlap integrals. In the case of continues control, the sine term we disregarded in the treatment should be taken into account like is done in [1].

The model for the noise in the control, modeled by $\Omega(t) = \Omega_0(t)\left[1 + n_c(t)\right]$, can be considered as general 1$^{\text{st}}$-order treatment of the noise. In this model the noise is gated (multiplied) by the function $\Omega_0(t)$. However, one can consider also the case of additive noise which is not gated by this function:

$$\Omega(t) = \Omega_0(t)\left[1 + n_c(t)\right] + b(t) \quad (12)$$

This noisy effect is the case if the control is derived from an oscillator which is controlled by an analogical gate which is noisy. However the effect of this noise, $b(t)$ has the same effect as $n_c(t)$, but gated with a function which is constant during the entire experiment time, $T$. To regard this in our treatment the filter function must be replaced by:

$$F_c(f) = \left| \int_0^T dt e^{-2\pi i f t} \right|^2 \quad , \quad (13)$$

which is just the original filter function plugging $\Omega_0 = 1$. In the case where the two kind of noises appear simultaneously, one should add an additional overlap integral. (or two additional overlap integrals in the case the noises $b(t)$ and $n_c(t)$ are correlated.)

## SHOT-TO-SHOT FLUCTUATIONS AND THE SYSTEM DENSITY OPERATOR

The evolution of the density matrix, described in our framework, is a sufficient description also for the fluctuation of the experimental results. For example, a strong noise in the control field can completely randomize the final state of the system. Consequently, the parameters $\rho_x$, $\rho_y$, and $\rho_z$, which parameterize the density operator, have to be zero, for this randomized state. Similarly, measuring the shot-to-shot fluctuation, one can calculate the average parameters: $\rho_x$, $\rho_y$, and $\rho_z$. It is important to note that unlike the $\sigma_z$ noise, the random effect of the control noise does not average out on the ensemble since all atoms are subjected to the same noise in a single realization. Therefore, more than a single experimental measurement has to be done to know the parameters $\rho_x$, $\rho_y$, and $\rho_z$.

The density operator fluctuations, are related to the average change in the density operator. This relation is captured by a geometrical interpretation of summing random vectors: Taking the sum of random vectors pointing in random direction, one gets a vector which length is smaller than the sum of all vector length. Describing the density operator by a vector and applying this argument, one can derive the relation between its mean value and fluctuations. However, we here derive it similarly to the previous section, starting from the evolution of the density operator. The master equation for the density operator up to the 1$^\text{st}$ order in the control noise is:

$$\dot{\tilde{\rho}}_c(t) = \frac{1}{i\hbar}\left[\tilde{H}_{int}(t),\, \tilde{\rho}(t)\right] \quad . \tag{14}$$

Plugging the interaction Hamiltonian, integrating over the short time $T$ and taking Trace to find the y-component of the vector, we obtain:

$$\Delta\rho_y(T) = Tr\{\Delta\rho_c(T)\sigma_y\} = Tr\left\{\frac{1}{i\hbar}\int_0^T dt n_c(t)\left[S_c(t),\rho(t)\right]\sigma_y\right\}$$

$$= \int_0^T dt\, n_c(t)\rho_z\Omega_0(t) \quad . \tag{15}$$

Taking the expectation value of $\Delta\rho_y(T)$ yields zero as the noise $n_c(t)$ averages to zero. However the fluctuations in this component is readily calculated by:

$$\left\langle\left(\Delta\rho_y(T)\right)^2\right\rangle = \left\langle\int_0^T dt_1\, n_c(t_1)\rho_z\Omega_0(t_1)\cdot\int_0^T dt_2\, n_c(t_2)\rho_z\Omega_0(t_2)\right\rangle \quad , \tag{16}$$

which in the frequency domain is written:

$$\left\langle\left(\Delta\rho_y(T)\right)^2\right\rangle = \rho_z^2\int_{-\infty}^{\infty} df G_c(f)F_c(f) \quad , \tag{17}$$

where the density operator change, $\Delta\rho_y(T)$, can be replaced by the component, $\rho_y(T)$ itself in this expression, since the initial density operator is not fluctuating.

## SHORT-TIMES FIDELITY

The fidelity $\mathcal{F} = Tr\{\rho(t)\rho(0)\}$ is derived for short times for an initial pure state. For pure dephasing, it is written by:

$$\mathcal{F} = Tr\{\rho(t)\rho(0)\} = Tr\{\left(\rho(0)+\Delta\rho(T)\right)\cdot\rho(0)\} \quad . \tag{18}$$

Since $\rho(0)$ is a pure state, the trace of $\rho^2(0)$ is 1 and therefore:

$$\mathcal{F} = 1 + Tr\{\Delta\rho(T)\cdot\rho(0)\}$$

$$= 1 + Tr\left\{\left(-\frac{\rho_x\sigma_x+\rho_y\sigma_y}{2}\frac{T}{2T}\int_{-\infty}^{\infty} dfG_\delta(f)F_\delta(f)\right)\frac{1}{2}\left(\rho_x\sigma_x+\rho_y\sigma_y+\rho_z\sigma_z+\mathbf{1}\right)\right\}$$

$$= 1 - \frac{1}{4}Tr\{\sigma_x^2\}\rho_x^2\frac{T}{2T}\int_{-\infty}^{\infty} dfG_\delta(f)F_\delta(f) - \frac{1}{4}Tr\{\sigma_y^2\}\rho_y^2\frac{T}{2T}\int_{-\infty}^{\infty} dfG_\delta(f)F_\delta(f)$$

$$= 1 - \frac{\rho_x^2}{4} \int_{-\infty}^{\infty} df G_\delta(f) F_\delta(f) - \frac{\rho_y^2}{4} \int_{-\infty}^{\infty} df G_\delta(f) F_\delta(f) \ . \tag{19}$$

Doing the same arithmetics with the noise in the control, one gets the fidelity of the system subjected to the two noises. It is written as:

$$\mathcal{F} = 1 - \frac{\rho_x^2 + \rho_y^2}{2} T R_\delta(T) - \frac{\rho_y^2 + \rho_z^2}{2} T R_c(T) \tag{20}$$

with the definitions for the decay rate functions:

$$R_\delta(T) = \frac{1}{2T} \int_{-\infty}^{\infty} df G_\delta(f) F_\delta(f) \ , \tag{21}$$

and

$$R_c(T) = \frac{1}{2T} \int_{-\infty}^{\infty} df G_c(f) F_c(f) \ . \tag{22}$$

The definitions of $R_\delta(T)$ and $R_c(T)$ are useful for since the functions are almost time independent for periodic pulses. This is due to the scaling of the filter function canceling the $1/T$ dependence. Hence the fidelity decays in a constant rate. It also helpful to note that the integral of filter function $F_\delta(f)$ is just T and theretofore, doesn't depend on the pulse frequency, but only on the total experiment time. However, the filter function $F_c(f)$ scales linearly with the pulse rate. Hence, the decay rate of the fidelity is large for spectral features at higher frequencies which is consistent with the experimental results.

---

[1] I. Almog, Y. Sagi, G. Gordon, G. Bensky, G. Kurizki, and N. Davidson, J. Phys. B **44**, 154006 (2011).


\end{widetext}

%\includepdf[pages={1}]{supplementary_material.pdf}
%\includepdf[pages={2}]{supplementary_material.pdf}
%\includepdf[pages={3}]{supplementary_material.pdf}
%\includepdf[pages={4}]{supplementary_material.pdf}
%\includepdf[pages={5}]{supplementary_material.pdf}

\end{document}